\documentclass[twocolumn,aps,prb,superscriptaddress, amsmath,amssymb,longbibliography]{revtex4-2}
\newcommand{\figurewidth}{.8\columnwidth}

\usepackage[utf8]{inputenc}
\usepackage[T1]{fontenc}

\usepackage{graphicx}
\usepackage{siunitx}	
\usepackage{amssymb}
\usepackage{datetime}
\usepackage{booktabs}
\usepackage{hyperref}
\usepackage{xcolor}
\hypersetup{
    colorlinks,
    linkcolor={red!50!black},
    citecolor={blue!50!black},
    urlcolor={blue!80!black}
}

\usepackage{todonotes}

\newcommand{\CdAs}{\texorpdfstring{Cd$_3$As$_2$}{Cd3As2}}
\newcommand{\Pc}{\ensuremath{P_{\text c}}}
\newcommand{\Pcu}{\ensuremath{P_{\text c \uparrow}}}
\newcommand{\Pcd}{\ensuremath{P_{\text c \downarrow}}}

\newcommand{\GPa}{\giga\pascal}
\newcommand{\RH}{\ensuremath{R_{\text H}}}
\newcommand{\Rt}{\ensuremath{R_{\text t}}}

\newcommand{\dd}{\text{d}}										

\begin{document}

%
%

\preprint{ver 0.7 \today\ \currenttime}

\title{Pressure-induced reconstructive phase transition in \CdAs\ }

\author{Monika Gam\.za}
\affiliation{Jeremiah Horrocks Institute for Mathematics, Physics and Astrophysics, University of Central Lancashire, Preston PR1 2HE, UK}

\author{Paolo Abrami}
\affiliation{HH Wills Laboratory, University of Bristol, Bristol, BS8 1TL, UK}

\author{Lawrence V D Gammond}
\affiliation{HH Wills Laboratory, University of Bristol, Bristol, BS8 1TL, UK}

\author{Jake Ayres}
\affiliation{HH Wills Laboratory, University of Bristol, Bristol, BS8 1TL, UK}

\author{Israel Osmond}
\affiliation{HH Wills Laboratory, University of Bristol, Bristol, BS8 1TL, UK}

\author{Takaki Muramtsu}
\affiliation{HH Wills Laboratory, University of Bristol, Bristol, BS8 1TL, UK}

\author{Robert Armstrong}
\affiliation{HH Wills Laboratory, University of Bristol, Bristol, BS8 1TL, UK}

\author{Hugh Perryman}
\affiliation{HH Wills Laboratory, University of Bristol, Bristol, BS8 1TL, UK}

\author{Dominik Daisenberger}
\affiliation{Diamond Light Source, Harwell Science and Innovation Campus, Didcot, OX11 ODE UK}

\author{Sitikantha Das}
\affiliation{HH Wills Laboratory, University of Bristol, Bristol, BS8 1TL, UK}
\affiliation{Department of Physics, Indian Institute of Technology, Kharagpur, 721 302, IN}

\author{Sven Friedemann}
\affiliation{HH Wills Laboratory, University of Bristol, Bristol, BS8 1TL, UK}
\email{Sven.Friedemann@bristol.ac.uk}

\date{\today}



\begin{abstract}
Cadmium arsenide \CdAs\ hosts massless Dirac electrons in its ambient-conditions tetragonal phase. 
We report X-ray diffraction and electrical resistivity measurements of \CdAs\ upon cycling pressure beyond the critical pressure of the tetragonal phase and back to ambient conditions. We find that at room temperature the transition between the low- and high-pressure phases results in large microstrain and reduced crystallite size both on rising and falling pressure. This leads to non-reversible electronic properties including self-doping associated with defects and a reduction of the electron mobility by an order of magnitude due to increased scattering. Our study indicates that the structural transformation is sluggish and shows a sizable hysteresis of over 1~GPa. Therefore, we conclude that the transition is first-order reconstructive, with chemical bonds being broken and rearranged in the high-pressure phase. 
Using the diffraction measurements we demonstrate that annealing at $\sim$\SI{200}{\celsius} greatly improves the crystallinity of the high-pressure phase. We show that its Bragg peaks can be indexed as a primitive orthorhombic lattice with $a_{\mathrm{HP}}~\approx$~8.68~\r{A}, $b_{\mathrm{HP}}~\approx$~17.15~\r{A} and $c_{\mathrm{HP}}~\approx$~18.58~\r{A}. 
The diffraction study indicates that during the structural transformation a new phase with another primitive orthorhombic structure may be also stabilized by deviatoric stress, providing an additional venue for tuning the unconventional electronic states in \CdAs.
\end{abstract}

\maketitle

\section{Introduction}
Electronic states with topological properties have been observed in a number of materials in recent years \cite{Armitage2018}. Among these, \CdAs\ is particularly important as it is stable in air and is easily synthesised via standard methods. The ambient-conditions tetragonal phase of \CdAs\ hosts gapless electronic states with a linear dispersion relation caused by symmetry-protected band crossings along $\Gamma-Z$ in the Brillouin zone, as predicted by Wang et al. \cite{Wang2013c} and subsequently observed by  ARPES and transport studies \cite{Liu2014b, Borisenko2014,Neupane2014,He2014}. The topological protection leads to Fermi arcs on the surface of \CdAs\ crystals \cite{Moll2016} and quantum Hall states along the edges \cite{Zhang2017c,Zhang2019a}. Recently, topologically protected states have also been observed in thin films of \CdAs\ \cite{Uchida2017,Schumann2018}. Thus, \CdAs\ is a prime candidate for technological applications based on topological properties, e.g. for optical switches, thermoelectrics, or based on the large magnetoresistance and possible topological superconductivity \cite{Zhu2017,Fu2020,Liang2014,Narayanan2015,Davydov2020}. Understanding the stability and formation of the ambient-conditions phase of \CdAs\ is important for fundamental research and applications, e.g.\ for production of \CdAs\ with high mobility in bulk and thin film. 

The ambient-conditions phase of \CdAs\ (phase I) hosting the topologically protected electronic excitations has a distorted superstructure of the CaF$_2$ type \cite{Ali2014}. Since \CdAs\ is Cd-deficient compared to the ideal antifluorite stoichiometry, missing two out of eight Cd atoms needed to form a simple cube around the As, the superstructure can be envisioned as consisting of distorted Cd$_6\Box_2$ cubes (where $\Box$ denotes an empty vertex) that are ordered in a spiralling corkscrew fashion along an axis parallel to the c-direction of the tetragonal crystal lattice. The latest ambient-pressure diffraction study revealed that each corkscrew is surrounded by corkscrews of the opposite chirality, resulting in the centrosymmetric crystal structure with the space group $I4_1/acd$ \cite{Ali2014}. As a consequence of the centrosymmetric symmetry, no spin splitting is allowed at the symmetry-protected band crossings and a bulk Dirac point at the Fermi energy along $\Gamma-Z$ results, making the ambient-conditions phase of \CdAs\ a 3D analogue of graphene. Breaking various symmetries using tuning parameters such as temperature or pressure may drive this system into other nontrivial quantum states \cite{Yang2014,Liu2014b}.   

With increasing temperature, \CdAs\ undergoes three consecutive structural phase transitions. It transforms to a Zn$_3$As$_2$-type structure ($P4_2/nbc$; phase II) at $\sim$\SI{220}{\celsius} and to another primitive tetragonal phase of the Zn$_3$P$_2$-type  ($P4_2/nmc$; phase III) at $\sim$\SI{470}{\celsius} before entering the CaF$_2$-type face-centred cubic (fcc) structure ($Fm\bar{3}m$; phase IV) at $\sim$\SI{600}{\celsius} which remains stable to the melting temperature \cite{Pistorius1975,Pietraszko1973,Arushanov1980}. 
Crystal structures of the high-temperature phases can be viewed as antifluorite superstructures containing 128, 32 and 8 distorted Cd$_6\Box_2$ cubes per conventional unit cell, respectively, and the transitions between them are related to changes in distribution of Cd atoms among the tetrahedral voids \cite{Ali2014}. In the high temperature fcc phase the Cd ions are disordered \cite{Pietraszko1973}, whereas in all the lower-temperature variants the Cd atoms form ordered patterns and shift from the ideal antifluorite positions toward the empty vertices of Cd$_6\Box_2$ cubes. The transition between the Zn$_3$As$_2$- and Zn$_3$P$_2$-type phases at \SI{470}{\celsius} has all the signatures of a typical first-order structural change, it proceeds quickly and with a notable thermal hysteresis. In contrast, the transition at \SI{220}{\celsius} from the Zn$_3$As$_2$-type structure to the ambient-temperature phase progresses slowly 
and shows no distinct hysteresis \cite{Pietraszko1973}. Evidence for an intermediate phase with disordered crystal structure was found at the isomorphic transition in Zn$_3$As$_2$ \cite{Pietraszko1973}. 

At high pressures additional phases of \CdAs\ have been identified \cite{Pistorius1975,Arushanov1980}. At room temperature a high-pressure semiconducting phase V was reported with a trigonal, orthorhombic or monoclinic symmetry \cite{Banus1969,Pistorius1975,Zhang2015a,He2016} above $\Pc\approx\SI{2.3}{\GPa}$; this phase was mapped out to be stable up to \SI{550}{\celsius}. Above \SI{550}{\celsius} three more high-pressure phases were detected but have not yet been characterised. Raman and optical reflectivity measurements indicated the presence of a consecutive phase at pressures above \SI{9.5}{\GPa} and \SI{8}{\GPa}, respectively \cite{Gupta2017,Uykur2018} where superconductivity was observed \cite{He2016}. 

Although the presence of  the  \CdAs\ phase V at high pressures has been established, the correlation between the structural and electronic characteristics is still under debate. In particular, some studies show that the metal-to-semiconductor transition coincides with the structural phase transition from the ambient-conditions phase I to the high-pressure  phase V \cite{Zhang2015a}, whereas other reports indicate that the onset of semiconducting behaviour occurs at much lower pressures ($\sim$1.1~GPa) compared to the structural phase transition 
\cite{He2016}. Furthermore, recent magnetotransport measurements revealed a sudden shrinkage of the Fermi surface and an anomaly in the phase factor coincident with an anisotropic compression of the crystal lattice at $p\approx$~1.3~GPa  interpreted as an indication for opening of a band gap \cite{Zhang2017d}. 
Those inconsistent findings motivated us to investigate in detail the pressure-driven evolution of the crystal structure and transport properties of \CdAs\ between the ambient-conditions phase I and the high-pressure phase V by means of high-resolution powder X-ray diffraction, electrical resistivity and Hall effect measurements.

\section{Methods}

\subsection{Experimental Details}
\label{subsection:ExperimentalDetails}

Needle-shaped single crystals were grown from self-flux in Cd excess as described in \cite{Ali2014}. High-purity Cd and As were loaded into an alumina crucible and sealed inside a quartz ampule. The materials were heated to \SI{825}{\celsius} and slowly cooled to \SI{425}{\celsius} before being centrifuged to remove excess Cd. As-grown single crystals displaying clear facets were used for electrical transport study. For powder X-ray diffraction measurements, crystals were ground with a slight amount of glycerol to avoid dust. 

Electrical resistivity and Hall effect measurements at room temperature were conducted in piston-cylinder cells with glycerol as a pressure transmitting medium. The resistance of a manganin wire was used as a continuous pressure gauge with \textit{in-situ} calibration on the resistive transitions of bismuth I-II and II-III \cite{Eremets1996}. The load on the piston-cylinder cell was regulated with a home-build hydraulic set-up enabling slow pressure sweeps with virtually constant $\dd P / \dd T$ (see inset of  \autoref{fig:RvsP}). The resistance and Hall effect were measured using a standard 4-point technique with a Stanford-Research SIM921 ac resistance bridge. The Hall effect was measured at fixed pressures in a $\pm\SI{0.2}{\tesla}$ electromagnet giving rise to a transverse resistance \Rt\ linear in magnetic field $B$. The Hall coefficient \RH\ was extracted from the slope of linear fits to $\Rt(B)$. Sample dimensions were measured with an optical microscope and give rise to a systematic uncertainty of $\pm\SI{15}{\percent}$ in \RH.

\begin{figure}
\includegraphics[width=.7\columnwidth]{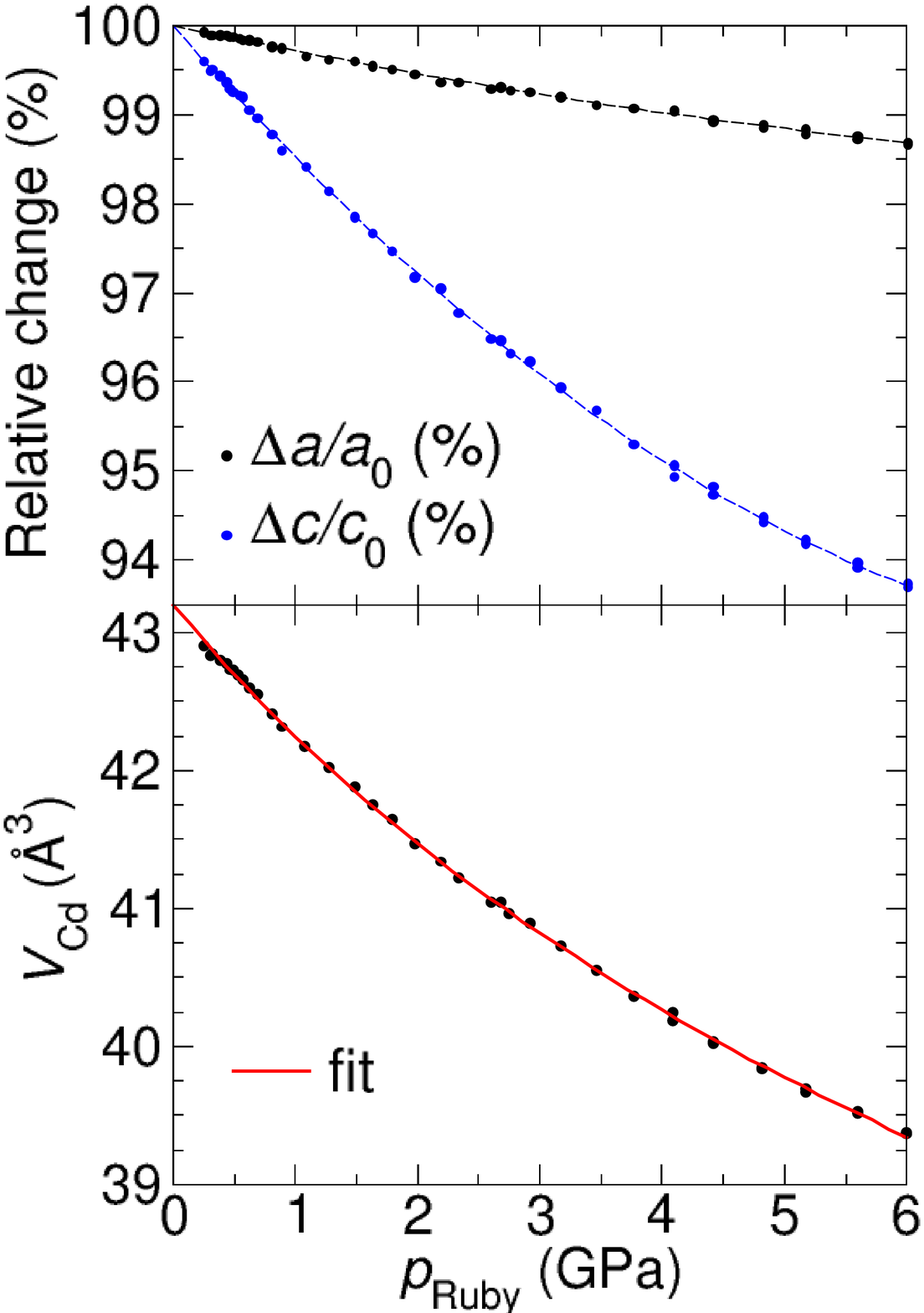}
\caption{Unit cell volume of Cd (bottom panel) and relative changes in its lattice parameters (top panel) obtained from Le Bail refinements of powder XRD patterns recorded on sample 1 at room temperature and plotted as a function of pressure measured using ruby fluorescence. Dashed lines are guides to the eye. Solid red line is the fit to the experimental data points using the Murnaghan equation of state with $K_0$~=~43~GPa, $K_0'$~=~8, and $V_0$~=~43.2~\r{A}$^3$, in agreement with \cite{EOS-Cd}. Error bars in the pressure values evaluated based on the deviations of the fit from the data points do not exceed 0.05~GPa.}
\label{fig:XRD-Cd}
\end{figure}

High-pressure powder X-ray diffraction (XRD) measurements where performed in membrane-driven diamond-anvil-cells (DACs) at the I15 beamline at Diamond Light Source.
X-ray radiation with wavelength  $\lambda = \SI{0.4246(5)}{\angstrom}$ was used throughout as calibrated on a LaB$_6$ standard. Pressure was measured using the ruby florescence method \cite{Syassen2008}. Diffraction peaks of residual cadmium flux mixed with \CdAs\ were used as a second pressure marker thus allowing to assess errors in the measured pressure values as shown in \autoref{fig:XRD-Cd}.
Three sample loadings were prepared and are labelled as sample 1, sample 2 and sample 3. For the first two loadings glycerol was used as a pressure medium whereas sample 3 was immersed in silicon oil. Temperatures up to \SI{250}{\celsius} have been accessed with an external heater for samples 2 and 3. Diffraction patterns were acquired on a MAR345 image plate detector and integrated with the DAWN software \cite{Filik2017} using the calibration parameters obtained from the LaB$_6$ standard. 

\subsection{Analysis of XRD patterns}
\label{subsection:XRDmethods}

Refinements of the diffraction patterns were carried out with Jana2006 \cite{Petricek2014a}. 
Profile refinements using Le Bail method were used for evaluating the structural parameters and for performing strain and size analyses. Contributions to the diffraction peak broadening originating in grain size and microstrain were separated based on their different dependencies on the scattering angle $\theta$. To this end, reflections visible in diffraction patterns in the 2$\theta$ range from 6$^{\circ}$ to $\sim$16$^{\circ}$ were fitted using Lorentzian and pseudo-Voigt shape functions. 
\footnote{The pseudo-Voigt function is an approximation of the Voigt function defined as:
\begin{equation}
pV(x) = \eta L(x) + (1-\eta) G(x),
\label{eq:1}
\end{equation}
with $L$($x$) and $G$($x$) denoting Lorentzian and Gaussian parts, respectively, that are of the same full width at half maximum (FWHM), $H$, and are weighted by $\eta$ adopting values between 0 and 1 and thus shifting the profile more towards pure Gaussian or pure Lorentzian. The $H$ and $\eta$ parameters are related to the FWHM values of the deconvoluted Lorentzian ($H_{\mathrm{L}}$) and Gaussian ($H_{\mathrm{G}}$) functions \cite{Thompson}: 
\begin{align*}
H &= H_{\mathrm{G}}^5 + 2.69269 H_{\mathrm{G}}^4H_{\mathrm{L}} + 2.42843 H_{\mathrm{G}}^3H_{\mathrm{L}}^2 
\\ 
&+ 4.47163 H_{\mathrm{G}}^2H_{\mathrm{L}}^3 
+ 0.07842 H_{\mathrm{G}}H_{\mathrm{L}}^4 + H_{\mathrm{L}}^5,
\\
\eta &= 1.36603\frac{H_{\mathrm{L}}}{H} - 0.47719 \left(\frac{H_{\mathrm{L}}}{H}\right)^2 
\\ 
&+ 0.11116\left(\frac{H_{\mathrm{L}}}{H}\right)^3.
\end{align*}}
The variations of the Gaussian and Lorentzian half widths with Bragg angle $\theta$ are described by the equations: 
\begin{equation}
H_{\mathrm{L}} = \frac{X}{\mathrm{cos}\theta} + Y \mathrm{tan}\theta + \Gamma_{\mathrm{L}}(\theta),
\label{eq:4}
\end{equation}
and 
\begin{equation}
H_{\mathrm{G}}^2 = 8ln2\left[ \frac{P}{\mathrm{cos}^2\theta} + U\mathrm{tan}^2\theta + \Gamma_{\mathrm{G}}^2(\theta) \right].
\label{eq:5}
\end{equation}
In each of \autoref{eq:4} and \autoref{eq:5}, the first term accounts for isotropic Scherrer particle broadening, the second term describes microstrain broadening related to lattice defects while the last term stands for the Lorentzian and Gaussian parts of instrumental broadening. The latter was assessed from powder XRD pattern of virtually strain-free LaB$_6$ powder with controlled grain size in the range of tens of micrometers. In the refinements, only one term describing grain size effect and one term describing microstrain broadening was used throughout.

According to the Scherrer formula \cite{Sch, Langford1978}, the volume weighted crystallite size, $D_{\mathrm{V}}$, is given by:
\begin{equation}
D_{\mathrm{V}} = \frac{K\lambda}{\beta \mathrm{cos}\theta},
\label{eq:6}
\end{equation}
where $K$ denotes the Scherrer constant and is assumed to be 1, $\lambda$ is wavelength of the radiation used for the XRD study, while $\beta$ represents integral breadth of a reflection located at 2$\theta$ expected assuming that the broadening originates solely in the grain size effect. \footnote{The integral breadths of the Lorentzian, Gaussian and normalized pseudo-Voigt functions are related to their FWHM as follows:
\begin{equation}
\beta_{\mathrm{L}} = \frac{\pi H_{\mathrm{L}}}{2},
\label{eq:12}
\end{equation}
\begin{equation}
\beta_{\mathrm{G}} = \frac{H_{\mathrm{G}}}{2} \sqrt{\frac{\pi}{\mathrm{ln}2}}, 
\label{eq:13}
\end{equation}
\begin{equation}
\beta_{\mathrm{pV}} = \frac{0.5\pi H}{\eta+(1-\eta)\sqrt{\pi\mathrm{ln}2}}, 
\label{eq:14}
\end{equation}
respectively.
}
Therefore, the volume averaged size of crystallites can be calculated from refined $X$ or $P$ values in degrees using the formulas:
\begin{align}
D_{\mathrm{V,L}} = \frac{360 K \lambda}{\pi^2X}    
\quad \text{and} \quad
D_{\mathrm{V,G}} = \frac{180 K \lambda}{\sqrt{2\pi^3P}}
\end{align}
for Lorentzian and Gaussian-type broadening attributed to the grain size effect, respectively.
The microstrain is given by 
\begin{equation}
\epsilon = \frac{\beta}{\mathrm{tan}\theta} .
\label{eq:9}
\end{equation}
Therefore, the respective Lorentzian and Gaussian contributions are determined as 
\begin{align}
\epsilon_{\mathrm{L}} = \frac{3\pi^2 Y}{1440} \quad \text{and} \quad
\epsilon_{\mathrm{G}} = \frac{\sqrt{2\pi^3 U}}{720} .
\label{eq:11}
\end{align}

XRD data for the high-pressure phase of \CdAs\ was also analysed by combining a Williamson-Hall method \cite{WH} with line profile analysis. To this end, a profile fitting procedure was adopted in which the measured XRD patterns were decomposed into individual peaks described using a pseudo-Voigt or Lorentzian shape function. According to the Williamson-Hall equation:
\begin{equation}
\beta \mathrm{cos}\theta = \frac{K\lambda}{D_{\mathrm{V}}} + 4\epsilon \mathrm{sin}\theta.
\label{eq:15}
\end{equation}
Therefore, the microstrain, $\epsilon$, and the volume weighted crystallite size, $D_{\mathrm{V}}$, are calculated from the slope and the intercept of the plot of  $\beta$cos$\theta$ against sin$\theta$ obtained after correcting integral breadths of the diffraction peaks for the instrumental broadening, assuming  $K$~=~1.          

Indexing of powder XRD patterns was performed using McMaille \cite{McMaille} and DISCVOL06 \cite{DISCVOL06} computer codes. Profile decomposition of the measured XRD patterns gave angular positions and intensities of individual peaks which were used as an input. Searching for candidate crystal lattices with volumes of up to 6000~\r{A}$^3$ was carried out in all  crystal systems by means of Monte-Carlo algorithms \cite{McMaille} or an exhaustive trial-and-error method with variation of parameters by successive dichotomy and partitioning of the unit cell volume \cite{DISCVOL06}. Space groups for the resulting sets of lattice parameters were identified using Jana2006 \cite{Petricek2014a}.

\section{Experimental Results}

We start by presenting continuous resistivity as well as discrete Hall effect measurements in \autoref{subsec:Res} showing the hysteresis and non-reversible behaviour of the electronic properties upon pressure-cycling of \CdAs\ into phase V and back to low pressure. Subsequently, we introduce XRD results in \autoref{subsec:XRD} which demonstrate the reversion to the ambient pressure tetragonal phase I and show poor crystallinity after the pressure cycling. Together these measurements provide evidence for a first order reconstructive phase transition. We evaluate the compressibility of the low- and high-pressure phases of \CdAs\ at room temperature. We show that annealing of the high-pressure phase at $\sim$\SI{200}{\celsius} greatly improves its crystallinity and thus allows to reveal more details about its crystal lattice. Finally, we indicate the formation of a new primitive orthorhombic phase prompted presumably by deviatoric stress upon releasing pressure through the structural phase transition in \CdAs.

\subsection{Electrical Transport}
\label{subsec:Res}

\begin{figure}%
\includegraphics[width = .7\columnwidth]{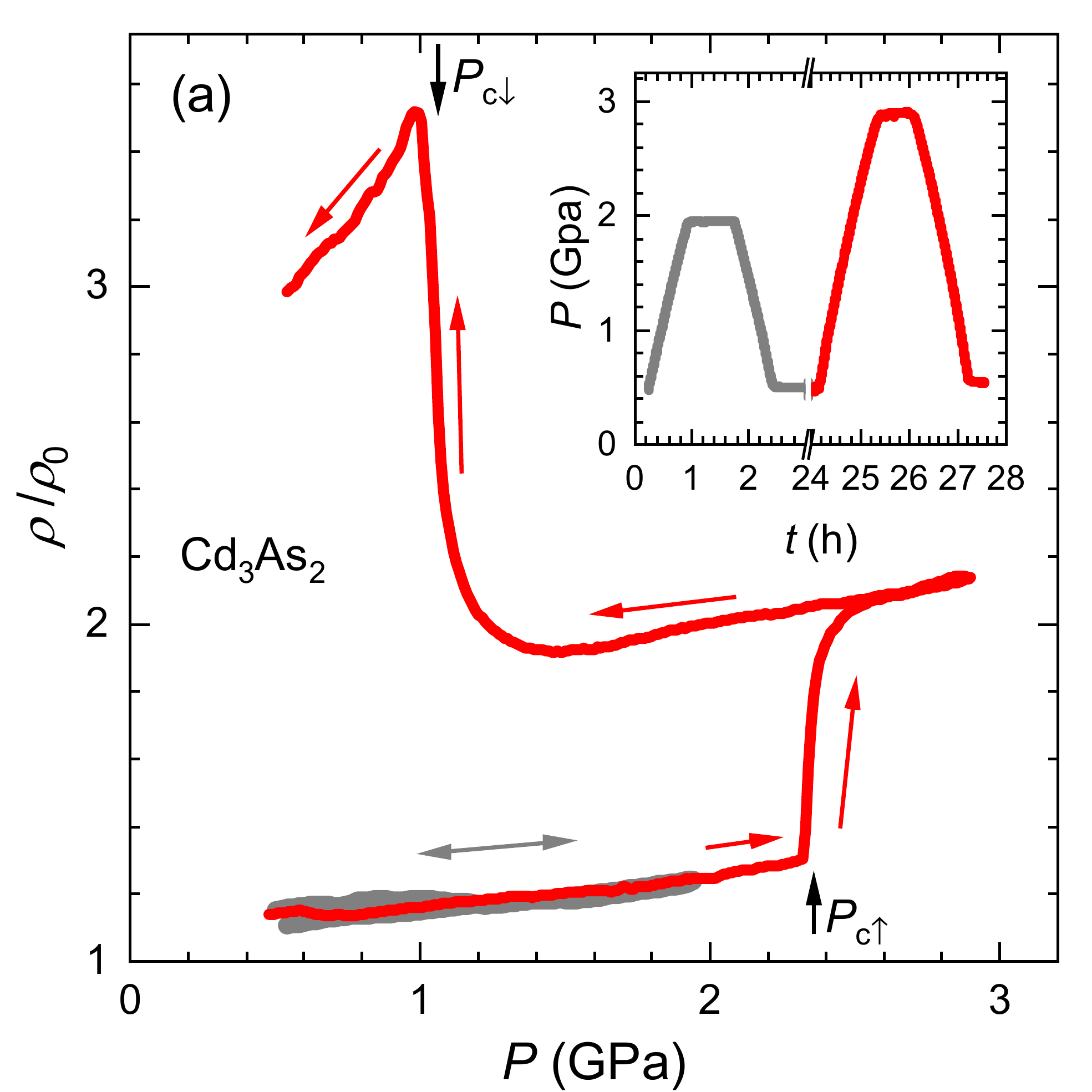}
\\
\includegraphics[width= .7\columnwidth]{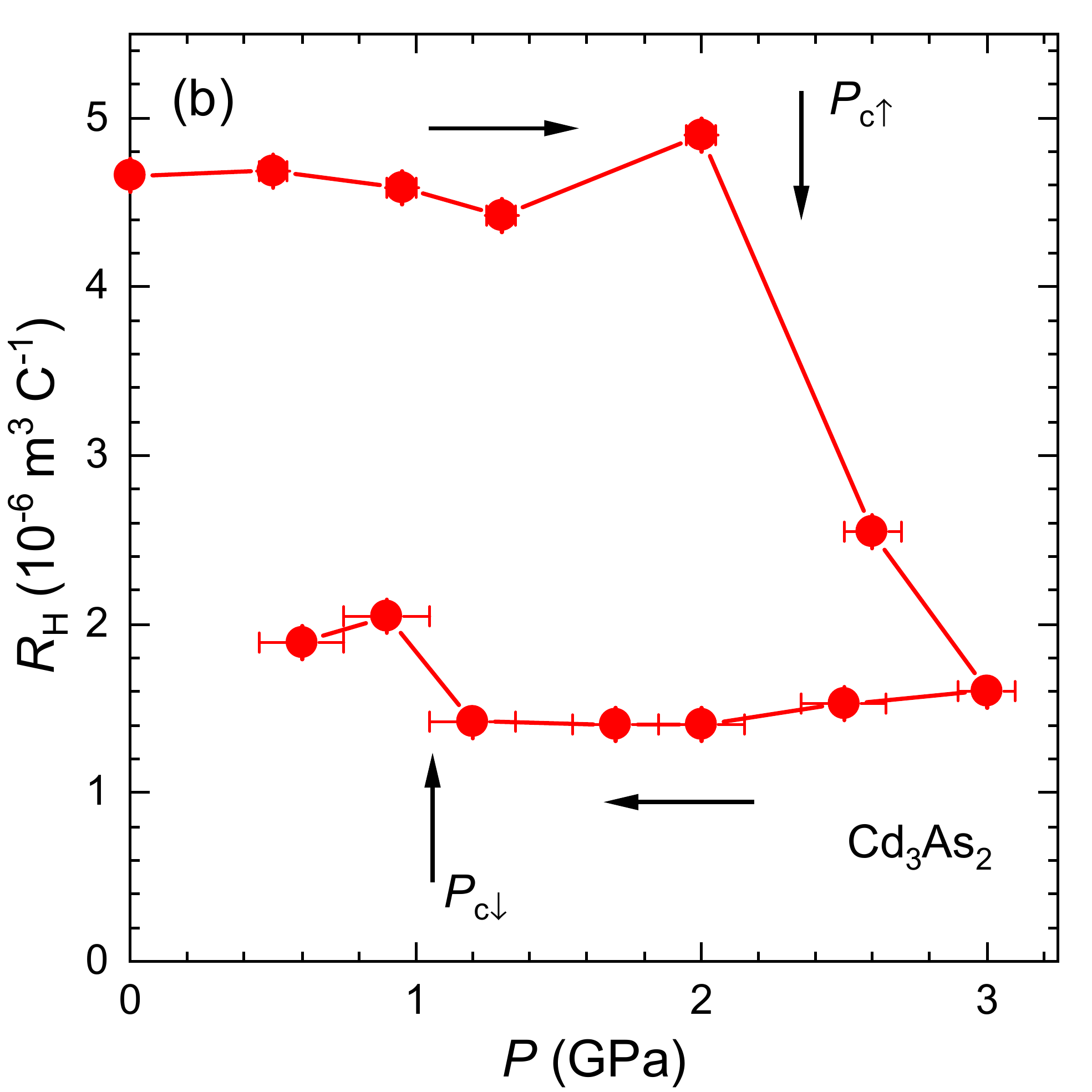}%
\caption{Non-reversible ambient temperature resistivity $\rho/\rho_0$ where $\rho_0 = \rho(P=0)$ (a) and Hall effect (b) under pressure in \CdAs. The initial pressure cycle from ambient pressure to \SI{2}{\GPa} carried out on a pristine sample is shown in grey. It was followed by the second pressure cycle to \SI{2.9}{\GPa} and back down to \SI{0.5}{\GPa} shown in red. The resistivity was normalised to the initial ambient pressure value $\rho_0$. Horizontal arrows indicate the evolution of pressure, vertical arrows highlight the position of the critical pressure on increasing pressure (\Pcu) and decreasing pressure (\Pcd). Inset shows the pressure protocol for the two cycles.}%
\label{fig:RvsP}%
\end{figure}%

As a crosscheck, we first measured the resistance of \CdAs\ on compression and decompression whilst staying within phase I. Using a pristine single crystal, the pressure was first  increased  to \SI{2}{\GPa} and subsequently released back down to \SI{0.5}{\GPa}.
As shown by grey lines in \autoref{fig:RvsP}(a), $\rho(P)$ is fully reversible within phase I and increases gradually with pressure. 

Once cycling the pressure through the phase transition into the \CdAs\ V phase , the resistance of \CdAs\ shows large hysteresis and non-reversible behaviour.
Subsequent to the cycling at low pressures, the pressure has been ramped to \SI{3}{\GPa} and released back down to \SI{0.5}{\GPa} as shown by the red data in \autoref{fig:RvsP}(a). The transition from phase I to phase V is clearly seen as a pronounced step-like increase in the resistance with a sharp onset at $\Pcu=\SI{2.3}{\GPa}$. Above \SI{2.6}{\GPa}, the resistance follows again a linear increase with further increasing pressure. The small linear increase below \Pcu, the jump at \Pcu, and the steeper linear increase above \Pcu\ are in very good agreement with earlier studies \cite{Zhang2015a,He2016}. Upon releasing the pressure, the resistance follows this linear regime to well below \Pcu\ indicating a large hysteresis of the phase transition. Only at  $\Pcd = \SI{1}{\GPa}$, a large rise and sharp cusp is observed  indicating a phase transition. Most notably though, the resistance does not return to the low pressure value. We note that the linear behaviour observed below \Pcd\ after pressure cycling extrapolates to a value more than double the starting value.

Our Hall effect measurements 
indicate a large increase in charge carrier concentration in the high-pressure phase of \CdAs. The results shown in \autoref{fig:RvsP}(b) were obtained upon stepwise increasing the pressure to \SI{3}{\GPa} and subsequently releasing it down to \SI{0.5}{\GPa} at room temperature. During raising pressure, the Hall coefficient remains roughly constant at $\RH=\SI{4.5e-6}{\meter\cubed\per\coulomb}$ in qualitative agreement with previous measurements at \SI{2}{\kelvin} \cite{Zhang2017d}. We note that the systematic difference by a factor of 3 and 10 with earlier studies by He et al. \cite{He2014} and Zhang et al. \cite{Zhang2017d}, respectively is probably due to uncertainties in the sample thickness and/or different level of doping whilst the difference between our room temperature and the previous low-temperature measurements is small \cite{He2014}. The qualitative agreement continues at high pressures where above \Pcu, the Hall coefficient is reduced to $\RH=\SI{1.6}{\meter\cubed\per\coulomb}$.
In a simple single-band model the drop of \RH\ at \Pcu\ corresponds to a three-fold increase of the charge carrier concentration $n$ in the high-pressure phase. This is similar to the 6-fold increase of $n$ observed at \SI{2}{\kelvin} by Zhang et al. \cite{Zhang2017d}. Together with the increase in resistance above \Pcu, this suggests an order-of-magnitude reduction of the carrier mobility in the high-pressure phase as also observed by Zhang et al. \cite{Zhang2017d}.

The initial value of \RH\ is not recovered after the pressure cycling. On decreasing pressure, the Hall coefficient remains roughly constant down to \Pcd\ with only a small increase below \Pcd\ to $\RH=\SI{2}{\meter\cubed\per\coulomb}$  -- a value close to early measurements of polycrystalline samples \cite{Rosenberg1959}.  
The irreversible behaviour in $\rho(P)$ and $\RH(P)$ suggests either the transformation into a phase different than the original ambient-conditions phase I or fundamentally modified electronic behaviour like increased scattering and/or charge carrier doping.
To explore the origin of the non-reversibility in the resistivity and Hall coefficient we performed detailed powder XRD measurements.

\subsection{X-ray diffraction} 
\label{subsec:XRD}

\begin{figure}
\includegraphics[width=\figurewidth]{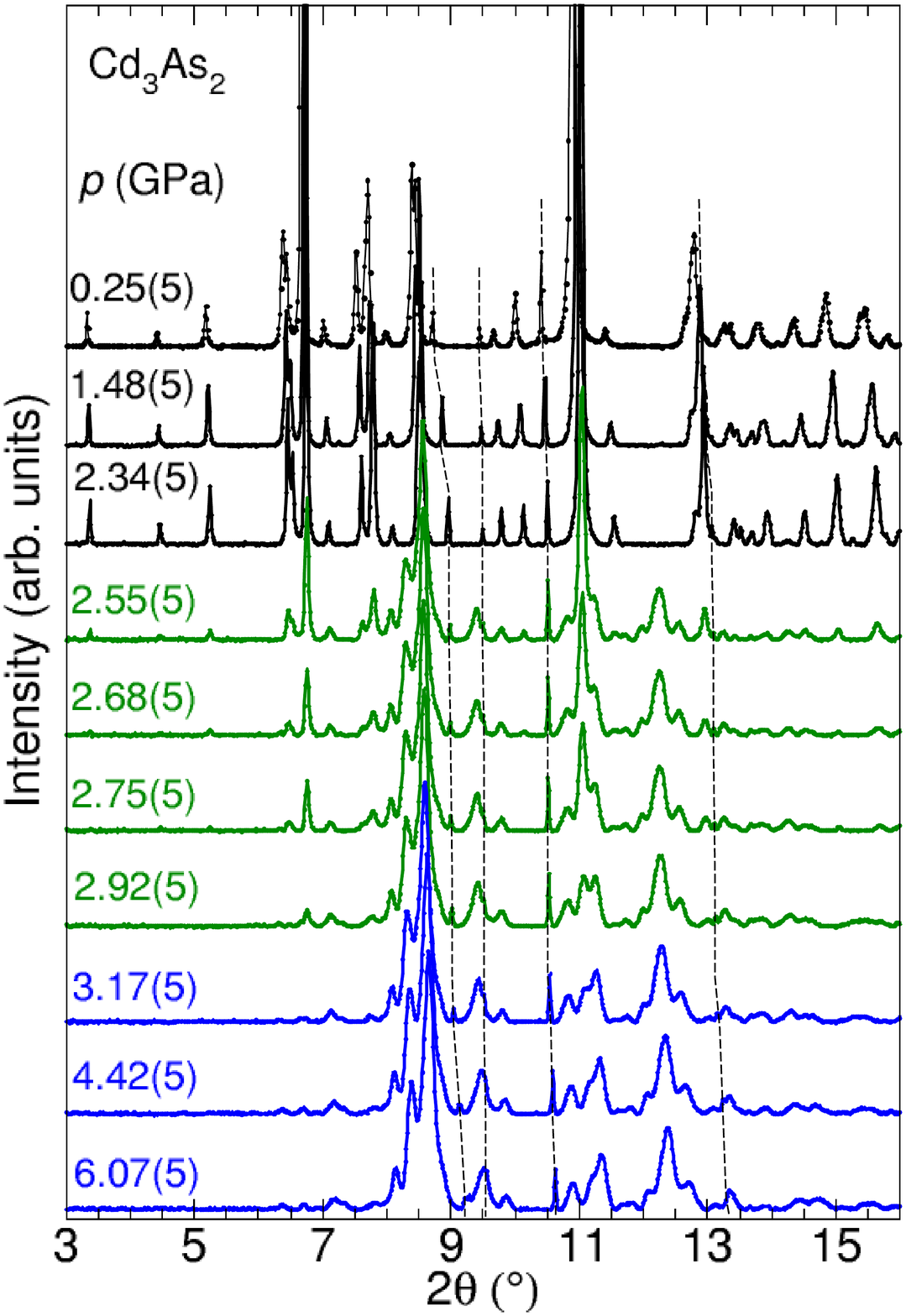}
\caption{Evolution of XRD pattern recorded for \CdAs\ at room temperature in sequence proceeding from top to bottom, i.e. pressure raised from \SI{0.25(5)}{\GPa} to \SI{6.07(5)}{\GPa}. Diffraction patterns consistent with the ambient-conditions phase I are shown in black, whilst blue colour is used for the XRD data corresponding to the high-pressure phase V. Diffraction patterns in green contain Bragg peaks originating in both the low- and high-pressure phases of \CdAs\ thus implying coexistence of phases I and V at pressures between 2.45(10) and $\sim$3~GPa. Dashed vertical lines indicate positions of peaks originating in Cd.}
\label{fig:XRD1}
\end{figure}

\begin{figure}
\includegraphics[width=.8\columnwidth]{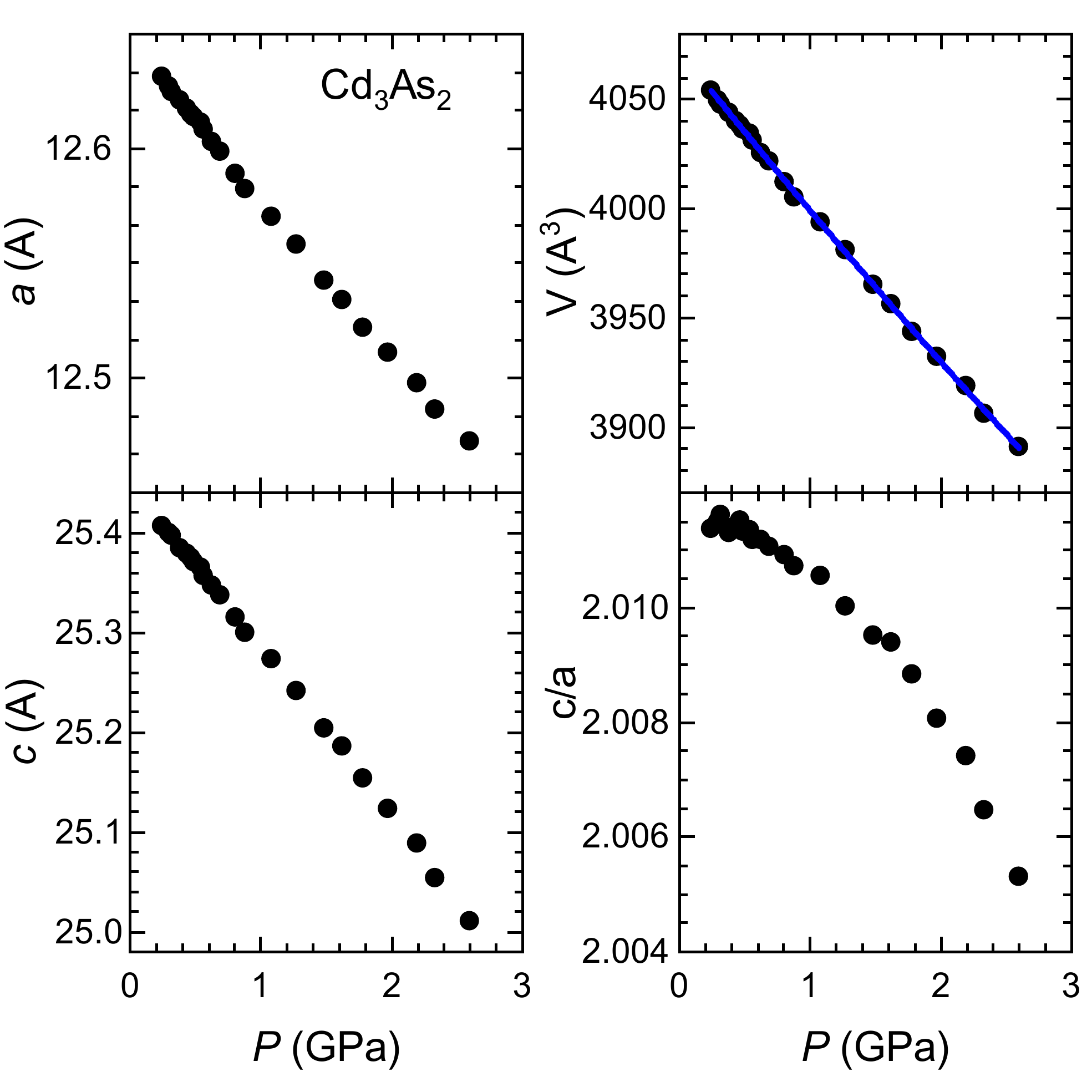}
\caption{Lattice parameters, their ratio and conventional unit cell volume of the ambient-conditions phase of \CdAs\ at room temperature plotted as a function of pressure. Error bars are smaller than the size of the circles.
Solid blue line is the fit to the experimental data points using the Murnaghan equation of state.}
\label{fig:lattice_param}
\end{figure}

Our XRD study on a pristine sample of the ambient-conditions phase of \CdAs\ after loading it into the pressure cell indicate that the powder was well-crystalline and single phased, except for a small contribution of cadmium flux detected  as  characteristic peaks at $2\theta\simeq\text{\SIlist{8.8;9.5;10.4;16.2}{\degree}}$ (see \autoref{fig:XRD1}) and used as a second pressure marker (see \autoref{fig:XRD-Cd}). To evaluate microstrain and sizes of \CdAs\ crystallites, Le Bail refinements of powder XRD patterns collected from several different sample spots were carried out. Excellent fits to the measured XRD data were achieved using pseudo-Voigt peak profiles with Gaussian contribution describing broadening due to the average grain size $D_{\mathrm{V}}$~=~170(30)~nm and with Lorentzian distribution accounting for microstrain $\epsilon$~=~0.24(2)\%. 
The rather significant positive microstrain indicates the presence of local compressive strain fields. They were likely created when dissipating mechanical energy during grinding single crystals of \CdAs\ into a fine powder suitable for the powder XRD study. The refined lattice parameters of the \CdAs\ phase I ($a$~=~12.6303(2)~\r{A}, $c$~=~25.409(1)~\r{A}, see \autoref{fig:lattice_param}) are very close to those reported in previous studies \cite{Ali2014, Steigmann1968, Zhang2017d}. 

\autoref{fig:XRD1} shows the evolution of XRD patterns collected under pressure at room temperature. With increasing pressure, the diffraction peaks are gradually shifted to larger diffraction angles, indicating a pressure-induced shrinkage of the unit cell.  For pressures up to \Pcu, the reflection pattern is consistent with the ambient-conditions tetragonal phase I of \CdAs. The widths and shapes of all the peaks remain virtually the same compared to immediately after closing the pressure cell indicating an elastic response of the crystal lattice to the applied stress field and giving no evidence for stress heterogeneity which would otherwise result in broadening of diffraction peaks. Lattice parameters derived from Le Bail refinements of the XRD patterns are shown in \autoref{fig:lattice_param}. The pressure-induced changes in the unit cell volume are well described by the Murnaghan equation-of-state fit resulting in $K_0$~=~54(1)~GPa, $K_0'$~=~2.3(8), and $V_0$~=~4072(1)~\r{A}$^3$. The bulk modulus of the ambient-conditions phase of \CdAs\ is smaller than that of copper and comparable to elemental Cd.

At pressures above $\sim$2.5~GPa, a clear alteration of the XRD pattern is observed (see \autoref{fig:XRD1}). This includes the loss of many peaks, e.g. for $2\theta\leq\SI{6}{\degree}$ and the emergence of new peaks incompatible with the space group $I4_1/acd$ of the phase I, e.g. at $2 \Theta \simeq \SI{12.5}{\degree}$, implying a change in the symmetry of the crystal lattice. The transition progresses slowly, with intensities of diffraction peaks originating in the low- and high-pressure phases changing gradually in the pressure range from $\sim$2.5 to $\sim$2.9~GPa and thus signifying a distinct phase coexistence regime.

The diffraction pattern of the high-pressure phase shown in \autoref{fig:XRD1} is similar to those reported in \cite{Zhang2015a, Zhang2017d, He2016} where it was interpreted as a monoclinic structure. All the diffraction peaks of the high-pressure phase appear very broad. At the same time, the small width of reflections originating from residual Cd flux highlights the high-quality pressure conditions with negligible pressure gradients. Thus, we conclude that the broadening of the \CdAs\ reflections is intrinsic to the sample. Since at low 2$\theta$ the effect of microstrain on widths of peaks is insignificant, such a huge broadening even at lowest diffraction angles indicates that during the pressure-induced structural phase transition the hundreds of nanometer-sized grains observed in the pristine  phase I form much smaller domains in the high-pressure \CdAs\ V phase, presumably due to local strains arising from enhanced crystalline defects. Estimates for the volume weighted domain size performed using \autoref{eq:6} based on integral breaths of the first four reflections of the high-pressure phase after correcting for the instrumental broadening gave $D_{\mathrm{V}}\approx$~16(6)~nm, an order of magnitude smaller than the average size of crystallites of the low-pressure phase before the transition.

\begin{figure*}
\includegraphics[width=\textwidth]{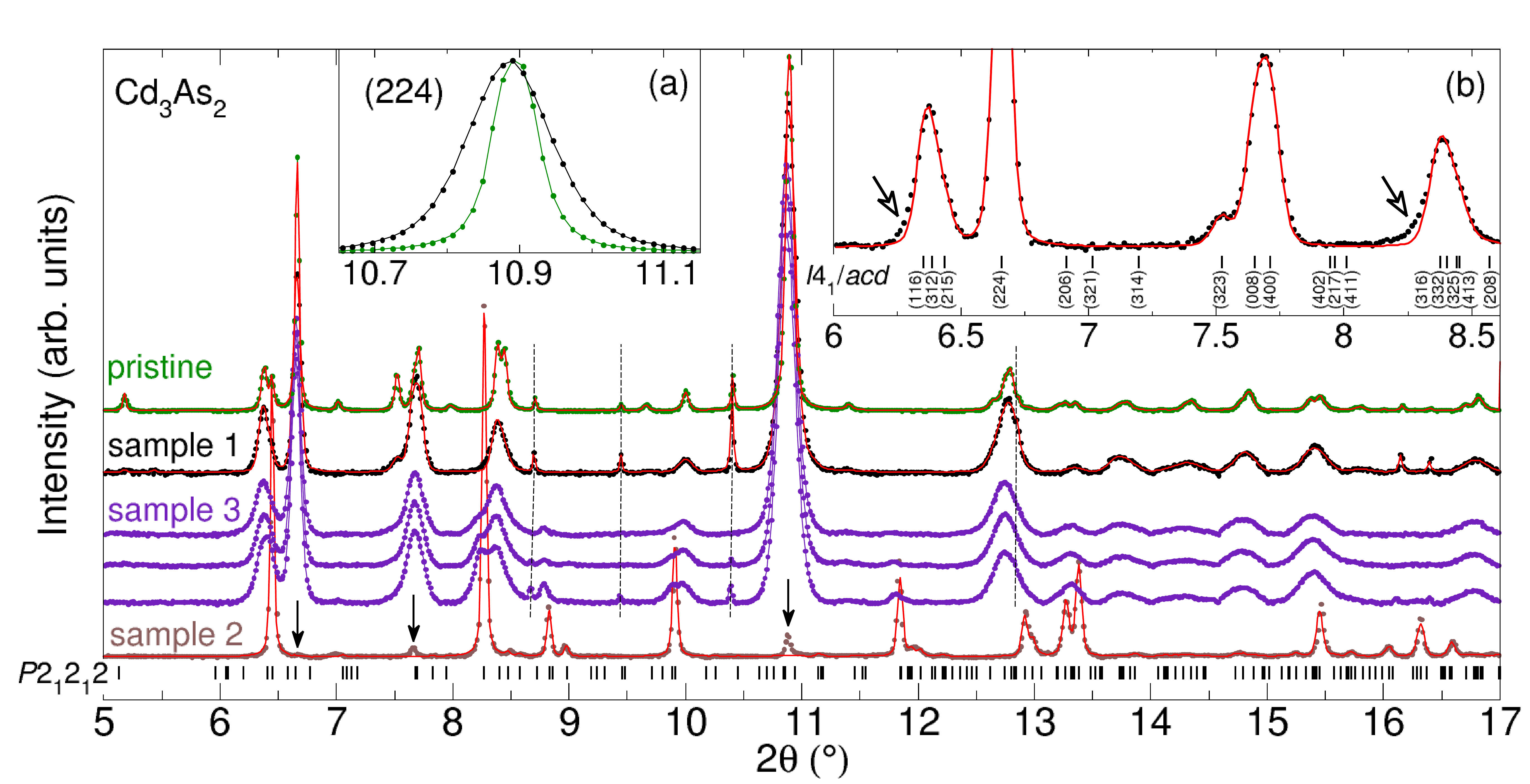}
\caption{Powder XRD patterns recorded at room temperature on \CdAs\ before (green dots, $P=\SI{0.25(5)}{\GPa}$) and after the pressure cycling. During the pressure cycling, the high-pressure phase was either kept at room temperature (sample 1 - black dots, $P=\SI{0.18(5)}{\GPa}$) or was annealed at temperatures up to \SI{240}{\celsius} and cooled down to room temperature before  pressure was released (sample 3 - purple dots, $P=0$; sample 2 - beige dots, $P=\SI{0.6(1)}{\GPa}$).  
Solid red line on top of the XRD data for sample 2 represents Le Bail refinement performed assuming pseudo-Voigt profiles for diffraction peaks consistent with an orthorhombic lattice with the space group $P2_12_12$ and the lattice parameters $a_{\mathrm{IP}}$~=~8.026(2)~\r{A}, $b_{\mathrm{IP}}$~=~4.1145(5)~\r{A} and $c_{\mathrm{IP}}$~=~18.988(1)~\r{A}. Angular positions of diffraction peaks expected for this orthorhombic lattice are indicated by short vertical black lines. 
Solid red lines on top of the XRD patterns for samples 1 and pristine correspond to Le Bail refinements performed assuming the ambient-conditions phase I crystal lattice. 
Vertical arrows mark peaks attributed to a small amount of the ambient-conditions phase in  sample 2. 
Dashed vertical lines indicate positions of diffraction peaks originating in Cd. 
Inset (a) shows the (224) reflection of the tetragonal $I4_1/acd$ phase I of \CdAs\ before and after the pressure cycling performed at room temperature normalized to the maximum intensity. Inset (b) shows enlarged angular range in which a pronounced asymmetric broadening is seen for some Bragg peaks of the ambient-conditions phase of \CdAs\ recovered after the room temperature pressure cycling. Short black lines in (b) indicate positions of reflections compatible with the crystal structure of the ambient-conditions phase.}
\label{fig:XRD3}
\end{figure*}

Upon release of pressure to below \SI{1.1}{\GPa}, the high-pressure V phase of \CdAs\  converts back into the ambient-conditions tetragonal structure. The phase transformation is complete at $p\approx$~0.7~GPa. Whilst, the peak positions of the tetragonal phase I have been recovered, the peak widths and shapes recorded after the pressure cycling are remarkably different, as shown in \autoref{fig:XRD3}. A strong perturbation of the relative peak intensities is evident especially in the range of $\SI{3}{\degree}\leq 2\theta \leq\SI{7}{\degree}$. Furthermore, the diffraction peaks of the ambient-pressure phase recovered after releasing pressure are much broader compared to the pristine sample. For instance, the (224) reflection shown in the inset of \autoref{fig:XRD3}(a) is broadened to nearly twice its original FWHM. Size and strain analyses based on Le Bail refinements of powder XRD patterns recorded after reducing pressure to 0.2--0.7~GPa indicated that the broadening of diffraction peaks from the low-pressure phase of \CdAs\ is dominated by microstrain. The best fits to the experimental data were achieved assuming Gaussian distribution describing the microstrain $\epsilon$~=~0.57(2)\% and a small Lorentzian or Gaussian contribution included to account for broadening due to the average coherent crystallite size $D_{\mathrm{V}}$~=~140(60)~nm similar to that before the pressure cycling. Noteworthy, a detailed inspection of the diffraction patterns measured after releasing pressure revealed that peak broadening is anisotropic and shows a distinct asymmetry, e.g. (116) and (316) peaks have pronounced tails on the low angle sides indicated by arrows in \autoref{fig:XRD3}(b). Both the asymmetry and $hkl$-dependence of peak profiles are typical for systems with a large degree of microstrain arising from high concentrations of dislocations and dislocation-type lattice defects \cite{M1, M2, M3, M4}.  In summary, the analysis of the peak widths provides evidence for large nonuniform strain and reduced crystallite domain size in the ambient-conditions phase I of \CdAs\ recovered after the pressure cycling.

\begin{figure*}
\includegraphics[width=\textwidth]{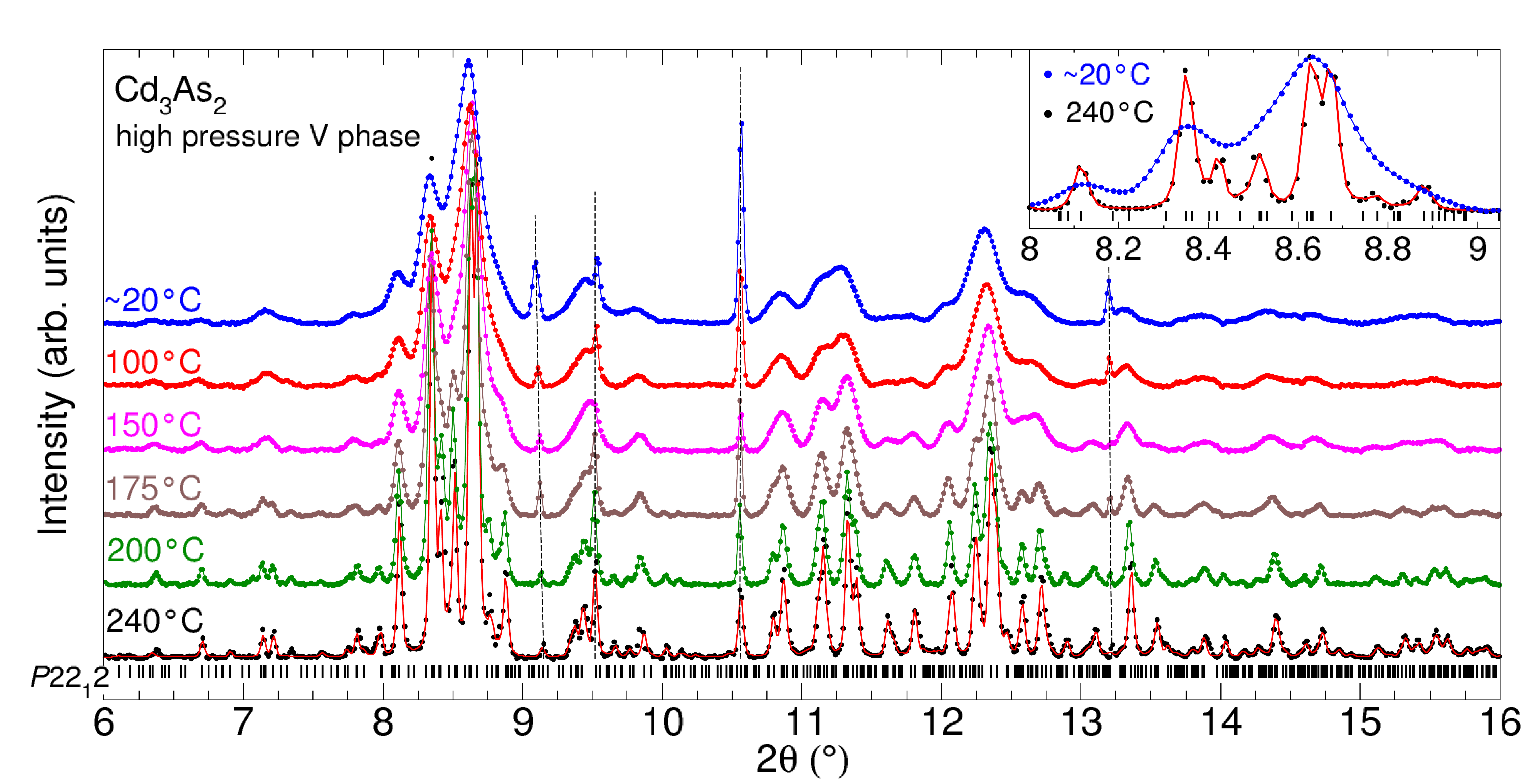}
\caption{X-ray diffraction patterns for the high-pressure  phase V of \CdAs\  measured on heating from room temperature to 240$^{\circ}$C during which the load was kept constant on the DAC but pressure has increased from \SI{4.18(5)}{\GPa} at room temperature to $\sim$\SI{6}{\GPa} at 240$^{\circ}$C. Thick solid red lines represent Le Bail refinement performed assuming pseudo-Voigt profiles for diffraction peaks consistent with an orthorhombic lattice with the space group $P22_12$ and the lattice parameters $a_{\mathrm{HP}}$~=~8.68~\r{A}, $b_{\mathrm{HP}}$~=~17.15~\r{A} and $c_{\mathrm{HP}}$~=~18.58~\r{A}. Angular positions of diffraction peaks expected for this orthorhombic lattice are indicated by short vertical black lines. Dashed vertical lines show positions of diffraction peaks originating in Cd. Inset shows enlarged angular range with the most intense peak manifold at room temperature (blue dots) and after annealing at 240$^{\circ}$C (black dots) normalized to the maximum intensity. The XRD pattern of the annealed sample revealed five intense Bragg peaks for 2$\theta$ between 8.2$^{\circ}$ and 8.7$^{\circ}$, whereas for the $P2_1/c$ structure proposed by Zhang et al. \cite{Zhang2015a} only two Bragg reflections are expected in this angular range.}
\label{fig:XRD_annealing}
\end{figure*}

The high-pressure phase of \CdAs\ can be annealed at temperatures between $\sim$\SI{150}{\celsius} and \SI{240}{\celsius}. The evolution of the XRD patterns collected a few minutes apart during heating is shown in \autoref{fig:XRD_annealing}. Upon increasing the temperature, the positions of diffraction peaks remain virtually unchanged due to a cancellation of thermal expansion and further compression as a result of the increasing pressure in the cell with friction locking the membrane. Most notably, raising the temperature above $\sim\SI{150}{\celsius}$ leads to a systematic narrowing of the diffraction peaks within minutes between measurements. The process of decreasing the peak widths accelerates with increasing temperature, as expected for a recrystallization anneal and grain growth \cite{4W}. It ends shortly after the temperature of \SI{200}{\celsius} is reached. The Williamson-Hall analyses (not shown) performed on several diffraction patterns recorded at temperatures between \SI{230}{\celsius} and \SI{240}{\celsius} on samples 2 and 3 indicated that after the annealing for both samples the microstrain was not larger than $\sim$0.06(\%) whereas the volume weighted domain size increased from $\sim$16(6)~nm before the heat treatment to $\sim$80(15)~nm after the annealing.

The crystal structure of the high-pressure phase of \CdAs\ remains elusive, but XRD patterns recorded on the annealed samples reveal more details about the symmetry of the crystal lattice. The previously suggested monoclinic lattice ($P2_1/c$, \cite{Zhang2015a}) is incompatible with the observed XRD patterns for the high-pressure V phase of \CdAs\ after the annealing. In particular, for the $P2_1/c$ structure proposed by Zhang et al. \cite{Zhang2015a} only two reflections are expected to contribute to the most intense peak manifold (see Fig.~1(d) in \cite{Zhang2015a}) whereas the diffraction patterns of the annealed samples revealed five intense Bragg peaks in the angular range of \SIrange{8.2}{8.7}{\degree}, as shown in \autoref{fig:XRD_annealing}. Further,  two Bragg peaks were detected at 2$\theta$~$\approx$~8.76$^{\circ}$ and 8.88$^{\circ}$ whilst no reflections are expected for the $P2_1/c$ structure in this angular range. 
The XRD patterns measured on samples 2 and 3 after the annealing at temperatures up to \SI{240}{\celsius} at $p$~=~5(1)~GPa were virtually identical. Bragg reflections detected consistently in those diffraction patterns were used for the purpose of indexing. Searching for candidate crystal lattices did not give satisfactory results for cubic, tetragonal, hexagonal and trigonal crystal systems with unit cell volumes of up to 6000~\r{A}$^3$. Among orthorhombic structures, the best indexing solution with the smallest unit cell volume that accounts for the observed diffraction peaks has a primitive lattice with $a_{\mathrm{HP}}~\approx$~8.68~\r{A}, $b_{\mathrm{HP}}~\approx$~17.15~\r{A} and $c_{\mathrm{HP}}~\approx$~18.58~\r{A}. We note that such a crystal lattice may be considered as a superstructure of the antifluorite structure type made of 96 distorted Cd$_6\Box_2$ cubes, similarly to the ambient-pressure polymorphs. 
Searching for monoclinic cells gave candidate lattices with comparable or only slightly smaller unit cell volumes ($\Delta V/V\lessapprox$~12\%). Notably, none of the lattices with smaller unit cell volumes could be interpreted in terms of the antifluorite-type building blocks. Therefore we conclude that the structure of the high-pressure phase of \CdAs\ is likely primitive orthorhombic. Space group tests indicated the chiral $P22_12$ or one of the lattices without systematic extinctions, i.e.\ centrosymmetric $Pmmm$, noncentrosymmetric chiral $P222$ or achiral $P2mm$, $Pm2m$, or $Pmm2$. Indexing of the high-pressure \CdAs\ V phase assuming the space group $P22_12$ is shown in \autoref{fig:XRD_annealing}.  
 
\begin{figure}
\includegraphics[width=.7\columnwidth]{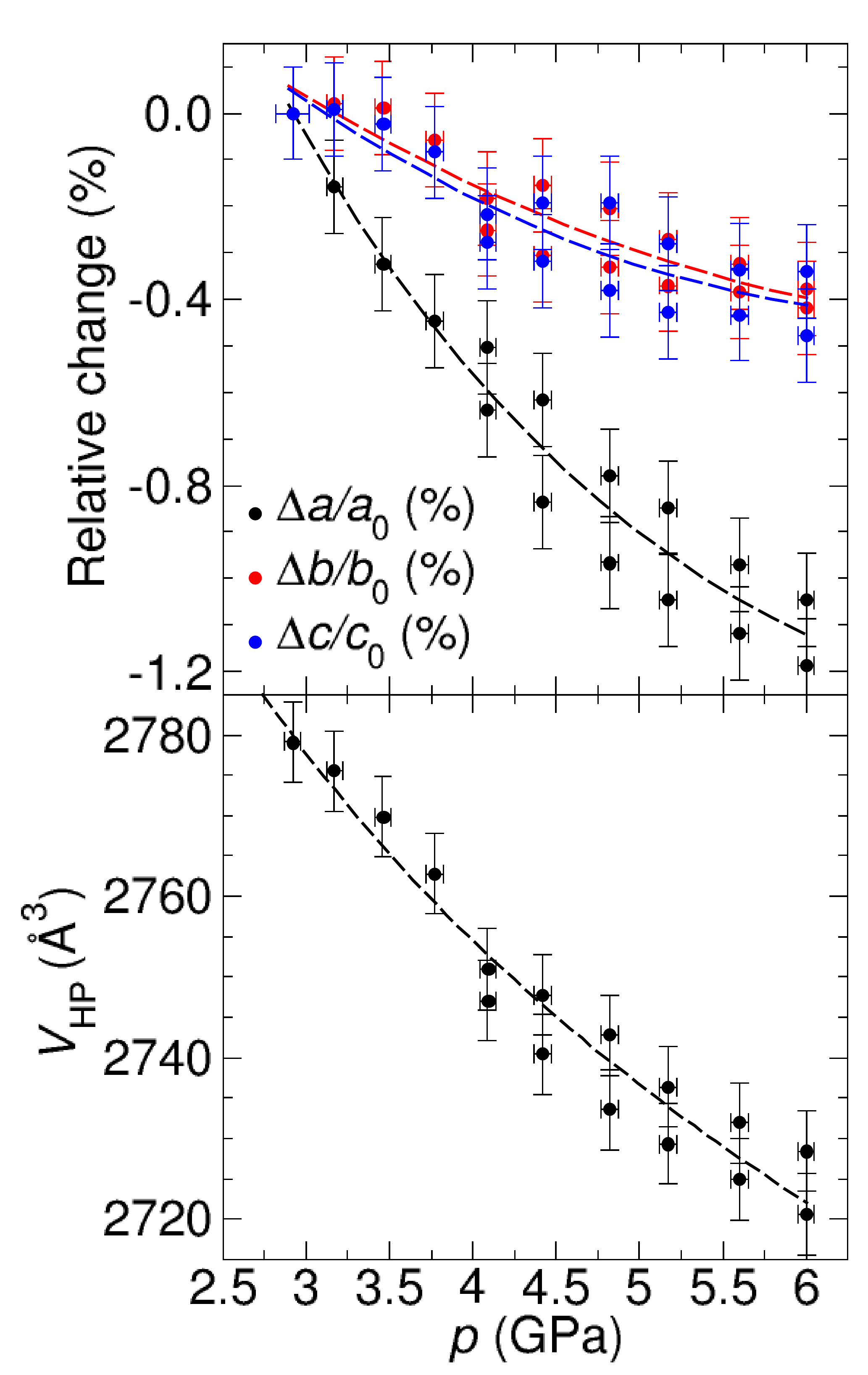}
\caption{The pressure dependence of the unit cell volume for the high-pressure phase V of \CdAs\ (bottom panel) and the pressure-induced relative changes in its lattice parameters (top panel) recorded at room temperature. Dashed lines are guides to the eye.}
\label{fig:XRD5}
\end{figure}

To evaluate the compressibility of the high-pressure phase of \CdAs\, sequential Le Bail refinements of powder XRD patterns collected at room temperature were performed assuming the orthorhombic crystal lattice. The resulting pressure dependencies of the lattice parameters and the unit cell volume are shown in \autoref{fig:XRD5}. The isothermal compressibility of the high-pressure phase immediately after the structural transformation is found to be $\kappa_{\mathrm{T}}$~=~0.9(1)$\times$~$10^{-11}$~Pa, \SI{45}{\percent} smaller than the compressibility of phase I ($\kappa_{\mathrm{T}}$~=~1.65(2)$\times$~$10^{-11}$~Pa, see \autoref{fig:lattice_param}), and it decreases further with pressure. 
The notable reduction in compressibility can be rationalized by an increase in the density of \CdAs\ at the structural transition by $\sim$4.8(2)\% estimated from the lattice parameters of the low- and high-pressure phases assuming that the latter is orthorhombic and consists of 96 distorted Cd$_6\Box_2$ cubes.

Upon lowering pressure below $\sim$1~GPa, the diffraction peaks of the annealed high-pressure phase in sample 3 start to diminish and finally the reflection pattern characteristic for the low-pressure phase of \CdAs\ emerges. As shown in \autoref{fig:XRD3}, the broadening of the diffraction peaks is very similar to that observed for sample 1 after the pressure cycling at room temperature. However, next to the peaks expected for the ambient-conditions phase of \CdAs\ there are also distinct additional contributions to the diffraction patterns at 2$\theta~\approx$~8.25$^{\circ}$, 8.78$^{\circ}$, 9.89$^{\circ}$, and 11.80$^{\circ}$ that may indicate the formation of a new phase. Whilst intensities of those extra peaks vary between different sample spots, their angular positions remain the same. We note that the additional features correspond well to the diffraction peaks observed for  sample 2 after annealing  at temperatures up to \SI{240}{\celsius} at pressures of 6~GPa and releasing the pressure. Since for  sample 2 glycerol was used as a pressure transmitting medium causing significant deviations from hydrostatic conditions in the accessed $T$-$p$ regime \cite{Tateiwa2009,Moulding2020}, we surmise that deviatoric stress induced by both the non-hydrostaticity of the pressure medium and larger size of the high-pressure phase crystallites can stabilize an intermediate metastable phase of \CdAs\ forming during the pressure-driven structural phase transition at room temperature. 

The diffraction peaks of the intermediate phase of \CdAs\ observed at ambient conditions can be indexed as a primitive orthorhombic structure with the lattice parameters $a_{\mathrm{IP}}$~=~8.026(2)~\r{A}, $b_{\mathrm{IP}}$~=~4.1145(5)~\r{A} and $c_{\mathrm{IP}}$~=~18.988(1)~\r{A} and the space group $P2_12_12$. We note that those lattice parameters are related to the lattice parameters of the high-pressure phase: $a_{\mathrm{IP}}~\approx~a_{\mathrm{HP}}$, $b_{\mathrm{IP}}~\approx~b_{\mathrm{HP}}$, and $c_{\mathrm{IP}}~\approx~c_{\mathrm{HP}}$/4, with the differences of +7.7\%, -2.2\%, and +3.9\%, respectively. Therefore, we suspect that the intermediate phase of \CdAs\ can be considered as another superstructure of the antifluorite structure type. Le Bail refinements performed assuming pseudo-Voigt profiles for Bragg peaks compatible with the orthorhombic lattice indicated a sizable degree of non-uniform strain, $\epsilon$~=~0.16(2)\%. The average crystallite size $D_{\mathrm{V}}$~=~108(10)~nm estimated from the refinements is comparable to the mean size of coherent domains of the high-pressure phase after annealing at temperatures of up to  \SI{240}{\celsius}.     

\section{Discussion}

Our high-resolution powder XRD measurement allowed us to evaluate in detail the changes in the lattice parameters of the ambient-conditions tetragonal phase I of \CdAs\ induced by the applied pressure.
While the observed pressure dependence of the unit cell volume shown in \autoref{fig:lattice_param} is similar to those reported in \cite{Zhang2017d, Zhang2015a}, our XRD data do not give any indication for a sudden jump in $c/a$ at pressures below \SI{1.5}{\GPa} that Zhang $et$ $al$. attributed to an additional topological phase transition \cite{Zhang2017d}. Instead, our  XRD study revealed a systematic decrease of the $c/a$ ratio with applied pressure. Remarkably, the observed changes in the $c/a$ ratio accord well with the pressure dependence of the Fermi wave vector derived from Shubnikov-de Haas oscillations measurements \cite{Zhang2017d}. 
Both the axial compression along the $c$ direction and the unexpected shrinking of the Fermi surface accelerate at  $p\gtrapprox$~\SI{1.7}{\GPa}. 
First principles electronic band structure calculations performed by Zhang $et$ $al.$ \cite{Zhang2017d} indicated that the shortening of the lattice parameter $c$ should shift the Dirac nodes towards the center of the Brillouin zone, thus reducing the Fermi wave vector and providing an explanation for the gradual increase of resistivity under pressure within the ambient-conditions phase of \CdAs\ (see \autoref{fig:RvsP}(a) and \cite{Zhang2015a}).

Our high-pressure studies provide a confirmation that the  I-V  transition in \CdAs\ is of first order type.
The electrical resistivity and powder XRD measurements both detect the phase transition with an onset at $\sim\SI{2.4}{\GPa}$ and at $\sim\SI{1.3}{\GPa}$ on raising and releasing pressure, respectively. The sizable hysteresis evidences the first order nature of the structural transformation which shows a significant width of $\sim\num{0.2}$ and $\sim\SI{0.5}{\GPa}$ at \Pcu\ and $\sim\num{0.4}$ and $\sim\SI{0.5}{\GPa}$ at \Pcd\ based on the electrical resistivity and XRD data, respectively. We note that some discrepancies between the two methods are not unusual because the former potentially probes a percolative path whereas the latter reflects changes in the entire sample volume. Importantly, the distinct pressure ranges with coexistence of the low- and high-pressure phases indicate sluggishness and thus an energy barrier between the phases I and V of \CdAs\. Therefore, we deduce that the transition process involves drastic changes in the system of chemical bonds and leads to a major reorganization of the crystal structure.

The observed increase in width of diffraction peaks 
at the transition from phase I to phase V indicates that the transformation causes fragmentation of crystallites into much smaller domains. Therefore, we attribute the tenfold reduction in mobilities of charge carriers inferred from the electrical transport study to the combined effects of an increased scattering resulting from the strongly reduced coherent domain size and the loss of Dirac electrons with high Fermi velocity caused by changes in the symmetry of the crystal lattice overcompensating the increase in charge carrier concentration.

Recovering the ambient-conditions phase I after cycling pressure at room temperature beyond \Pcu\ resulted in a large degree of microstrain. Estimates from profile refinements using the Le Bail method gave $\epsilon \approx \SI{0.57(2)}{\percent}$, which is three orders of magnitude larger than $\epsilon$ values $\sim10^{-6}$ expected for nearly strain-free crystals \cite{SF} and nearly three times larger than the microanstrain induced by grinding crystals into a fine powder with tens of a nanometer size grains. Although results of the size and strain analyses need to be considered with caution because of significant peak overlap, the large positive $\epsilon$ indicates the presence of local compressive stress fields arising from lattice disorder such as dislocations, stacking faults, intercalations of atoms or grain boundaries. Such defects may give rise to  unintentional doping \cite{doping, Hyuga_1985}, providing a plausible explanation for the observed decrease in the Hall coefficient for the ambient-conditions phase I of \CdAs\ after  the pressure cycling (see \autoref{fig:RvsP}(b)). Thus, we infer that the unexpected increase in resistivity after returning to the low pressure regime below \Pcd\ shown in \autoref{fig:RvsP}(a) originates in a largely enhanced scattering that outweighs the  expected increase in carrier mobilities due to the formation of bulk Dirac points and the gain in the charge carriers  from self-doping. We note that studies on \CdAs\ epilayers revealed a substantial effect of extended defects on electron mobilities \cite{Rice2019}. Moreover, for epitaxial (112) \CdAs\ films a systematic decrease in Hall mobilities with the compressive in-plane strain ($\epsilon_{\mathrm{s}}\lessapprox$~0.4\%) was observed \cite{ThinFilms1}. Further study including high resolution transmission electron microscopy imaging is needed to address the origin of the increased scattering in the ambient-conditions phase of \CdAs\ retrieved after the room-temperature pressure cycling.

\section{Conclusions}

Our high-pressure studies highlight the disruptive nature of the pressure-driven structural phase transition in \CdAs. We find large changes in  both the electronic and microstructural characteristics at the room-temperature structural transformation. The reduction of the carrier mobility is shown to result from the formation of microstrain and fragmentation of crystallites into smaller domains 
caused by the phase transition. Our study indicates that the increased resistance is dominated by the enhanced scattering outweighing the gain in charge carriers in the high-pressure phase V. We show that annealing of the high-pressure phase at temperatures of $\sim$\SI{200}{\celsius} results in a largely improved crystallinity and thus should enable future studies aiming at solving the crystal structure and exploring the electronic properties. Finally, we observe the formation of a new primitive orthorhombic phase prompted presumably by deviatoric stress upon releasing pressure through the structural phase transition. Our results highlight the importance of strain and crystallite size in tailoring electronic properties and demonstrate a route to annealing \CdAs\ for future applications of bulk and thin-film.

\begin{acknowledgments}
The authors would like to thank Ingo Loa for valuable discussion and Charles Clapham and Chris Bell for technical support. This work was partially supported by the EPSRC under grants EP/R011141/1, EP/L025736/1, EP/N026691/1 as well as the ERC Horizon 2020 programme under grant 715262-HPSuper. We acknowledge Diamond Light Source for time on beamline I15 under proposal EE19319-1.
\end{acknowledgments}

\section*{Additional information}

Data are available at the University of Bristol data repository, \href{https://data.bris.ac.uk/data/}{data.bris}, at \url{https://doi.org/10.5523/bris.xxxx} \cite{Cd3As2_data}.

\section*{References}
%


\end{document}